\documentclass[conference]{IEEEtran}
\IEEEoverridecommandlockouts
\usepackage[numbers,sort&compress]{natbib}
\usepackage{multirow}
\usepackage[flushleft]{threeparttable}
\usepackage{latexsym}
\usepackage{amsmath}
\usepackage{amssymb}
\usepackage{amsthm}
\usepackage{graphicx}
\usepackage{epsfig}
\usepackage{amssymb}
\usepackage{subcaption}
\usepackage{mwe}
\usepackage{float}
\usepackage{mathtools}
\usepackage{algorithm}
\usepackage[noend]{algpseudocode}
\usepackage[utf8]{inputenc}
\usepackage[english]{babel}
\usepackage{fancyhdr}
\usepackage{mleftright}
\usepackage{courier}
\usepackage{array}
\usepackage{booktabs}
\usepackage{syntonly}
\usepackage{xcolor}
\usepackage{colortbl}
\usepackage{lipsum}
\usepackage{epstopdf}
\usepackage{environ}
\usepackage{relsize}
\usepackage{multirow}
\usepackage{diagbox}
\usepackage{color} 
\usepackage[normalem]{ulem} 
\usepackage[makeroom]{cancel} 
\usepackage{bm}
\usepackage{enumitem}
\usepackage{makecell}
\usepackage[colorlinks,urlcolor=black,citecolor=black,linkcolor=black,anchorcolor=black]{hyperref}

\newcommand{\PreserveBackslash}[1]{\let\temp=\\#1\let\\=\temp}
\newcolumntype{C}[1]{>{\PreserveBackslash\centering}p{#1}}
\newcolumntype{R}[1]{>{\PreserveBackslash\raggedleft}p{#1}}
\newcolumntype{L}[1]{>{\PreserveBackslash\raggedright}p{#1}}
\makeatletter
\let\OldStatex\Statex
\renewcommand{\Statex}[1][3]{%
  \setlength\@tempdima{\algorithmicindent}%
  \OldStatex\hskip\dimexpr#1\@tempdima\relax}
\makeatother

\def\BibTeX{{\rm B\kern-.05em{\sc i\kern-.025em b}\kern-.08em
    T\kern-.1667em\lower.7ex\hbox{E}\kern-.125emX}}

\begin{document}

\title{An Optimization Method-Assisted Ensemble Deep Reinforcement Learning Algorithm to Solve Unit Commitment Problems} 

\author{Jingtao Qin,~\IEEEmembership{Student Member,~IEEE,} Yuanqi Gao,~\IEEEmembership{Member,~IEEE,} Mikhail Bragin,~\IEEEmembership{Member,~IEEE,} \\ and Nanpeng Yu,~\IEEEmembership{Senior Member,~IEEE,}
}

\markboth{Journal of \LaTeX\ Class Files,~Vol.~14, No.~8, August~2021}%
{Shell \MakeLowercase{\textit{et al.}}: A Sample Article Using IEEEtran.cls for IEEE Journals}


\maketitle

\begin{abstract}
Unit commitment (UC) is a fundamental problem in the day-ahead electricity market, and it is critical to solve UC problems efficiently. Mathematical optimization techniques like dynamic programming, Lagrangian relaxation, and mixed-integer quadratic programming (MIQP) are commonly adopted for UC problems. However, the calculation time of these methods increases at an exponential rate with the amount of generators and energy resources, which is still the main bottleneck in industry. Recent advances in artificial intelligence have demonstrated the capability of reinforcement learning (RL) to solve UC problems. Unfortunately, the existing research on solving UC problems with RL suffers from the curse of dimensionality when the size of UC problems grows. To deal with these problems, we propose an optimization method-assisted ensemble deep reinforcement learning algorithm, where UC problems are formulated as a Markov Decision Process (MDP) and solved by multi-step deep Q-learning in an ensemble framework. The proposed algorithm establishes a candidate action set by solving tailored optimization problems to ensure a relatively high performance and the satisfaction of operational constraints. Numerical studies on IEEE 118 and 300-bus systems show that our algorithm outperforms the baseline RL algorithm and MIQP. Furthermore, the proposed algorithm shows strong generalization capacity under unforeseen operational conditions.

\end{abstract}


\section{Introduction}
\IEEEPARstart{U}{nit} Commitment (UC) is a crucial decision-making tool used by Independent System Operators (ISOs) in the day-ahead electricity market. In UC problems, the optimal schedule of generators needs to be determined given the supply offers, demand bids, transmission network situations, and operational limits. The UC problems can be classified into different subgroups in a few ways \cite{abdou2018unit}. With respect to security constraints, UC problems can be divided into conventional UC problems and security-constrained UC (SCUC) problems \cite{fu2013modeling,wang2008security}. According to whether uncertainty is considered and whether the probabilistic distribution of uncertain parameters is known \cite{yang2021comprehensive}, UC can be categorized into deterministic UC problems, stochastic UC problems \cite{wang2012stochastic, zhao2015data}, and robust UC problems \cite{bertsimas2012adaptive, wang2022distributionally, zhang2020partition, chen2018distributionally}. 

 It is critical for enhancing the efficiency of the day-ahead electricity market to obtain near-optimal solutions to UC problems. The existing solution approaches to UC problems include heuristic algorithms \cite{wang2018extended,senjyu2006emerging}, mathematical optimization algorithms, intelligent optimization algorithms \cite{nemati2018optimization,ting2006novel,simopoulos2006unit}, and machine learning (ML) based approaches. Among these approaches, mathematical optimization algorithms including dynamic programming (DP) \cite{snyder1987dynamic}, branch-and-cut algorithm \cite{gao2022internally}, Lagrangian relaxation (LR) \cite{ongsakul2004unit}, Benders decomposition \cite{liu2010extended}, outer approximation \cite{yang2016multi}, ordinal optimization \cite{wu2013stochastic}, and column-and-constraint generation \cite{an2014exploring} have been widely studied in UC problems. Even though satisfactory performance is achieved by these methods, their calculation time grows at an exponential rate with the amount of energy resources. Thus, obtaining a near-optimal UC solution efficiently can be difficult when the renewable energy resources and corresponding uncertainties keep rising. To improve the performance of large-scale UC problem solvers, some researchers try to improve the tightness and compactness of the UC problem formulation as a MILP model \cite{morales2013tight,atakan2017state,yan2019systematic}. Besides, a decomposition and coordination approach is proposed in \cite{sun2018novel}, which leverages Surrogate Lagrangian Relaxation (SLR) to solve large-scale UC problems with near-optimal solutions within time limits. Furthermore, Surrogate Absolute-Value Lagrangian Relaxation (SAVLR) is enhanced in \cite{wu2021novel} by embedding the ordinal-optimization method to drastically reduce the solving time of subproblems. Recently, a novel quantum distributed model is proposed to solve large-scale UC problems in a decomposition and coordination-supported framework \cite{nikmehr2022quantum}. A temporal decomposition method was proposed in \cite{kim2018temporal} which systematically decouples the long-horizon MIP problem into several sub-horizon models.

The main limitation of mathematical optimization algorithms is that they assume one-shot optimization where the UC problems need to be solved from scratch each time. In practice, UC problems are solved on a daily basis in the day-ahead market with changes to the input data \cite{xavier2021learning} while the structure of the problem formulation stays the same. Thus, the previous UC problems' solutions provide useful information that can be utilized to improve the solution quality of similar UC problems.

The recent advances in artificial intelligence motivate the development of machine learning-based methods to solve UC problems \cite{yang2021machine}. A series of machine learning techniques are proposed to extract valuable information from solved instances of UC problems to enhance the warm-start capabilities of MIP solvers in \cite{xavier2021learning}. Neural networks are developed to imitate expert heuristics and speed up the branch-and-bound (B\&B) algorithm, which achieves significant improvements on large-scale real-world application datasets including Electric Gird Optimization \cite{nair2020solving}. Unlike supervised learning, which requires labeled data, reinforcement learning (RL) is a mathematical tool for learning to solve sequential decision-making problems such as volt-var control problems in power distribution systems \cite{wang2019safe}. In \cite{jasmin2009reinforcement}, UC problems for a system with 4 units are modeled as multi-stage decision-making tasks, and RL solutions are formulated through the pursuit method. Three RL algorithms including approximate policy iteration, tree search, and back sweep are proposed to minimize operational costs on a 12-unit system \cite{dalal2015reinforcement}. The UC problem with 10 units is tackled as a multi-agent fuzzy RL task, and units play as agents to corporately reduce the overall operation cost \cite{navin2019fuzzy}. A method based on decentralized Q-learning to find a solution to UC problems on a system with up to 10 units is introduced in \cite{li2019distributed}. An RL-based guided tree search algorithm is developed to solve stochastic UC problems for a system with 30 generation units \cite{de2021applying}, which uses a pre-trained policy to reduce the action space and designs a neural network as a binary classifier that sequentially predicts each bit in the action sequence. Most of the existing RL-based algorithms have only been tested on small-scale UC problems because they suffer from the curse of dimensionality. Specifically, the number of states and feasible actions increases exponentially with the size of the UC problems. Furthermore, many operational constraints such as the transmission line capacity limit can not be strictly enforced in these RL-based algorithms.

\begin{figure}[htbp]
\centerline{\includegraphics[width=9cm]{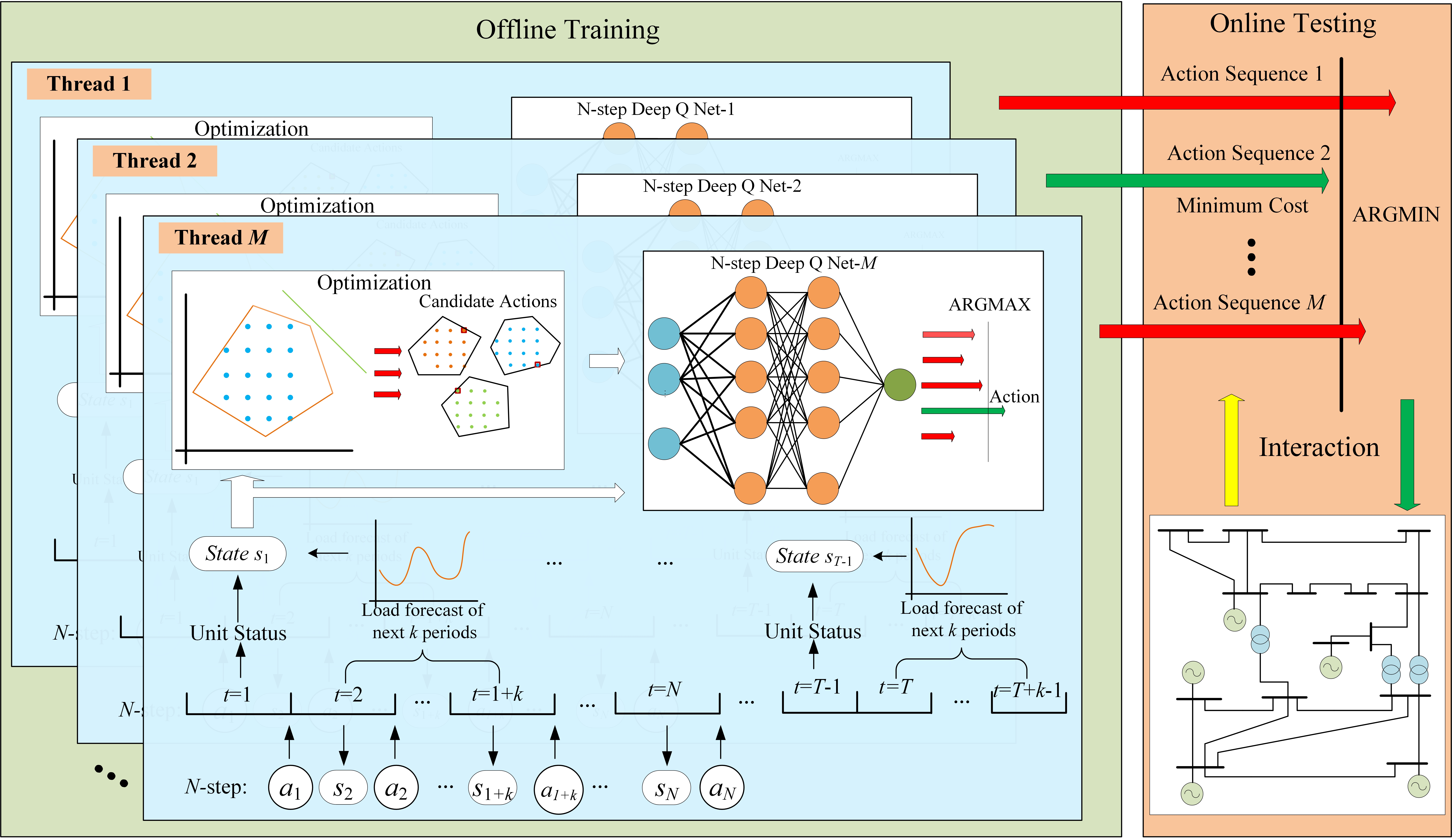}}
\caption{The overall framework of optimization method-assisted ensemble RL algorithm.}
\label{fig1}
\end{figure}

To address the limitations of the existing algorithms, we synergistically combine the mathematical optimization method with RL and propose an optimization method-assisted ensemble deep reinforcement learning algorithm to solve deterministic UC problems. The overall framework of the proposed approach is shown in Fig. \ref{fig1}. First, we establish a candidate action set by solving a series of simplified optimization problems to ensure that the solutions are feasible and can achieve decent performance. These candidate actions will serve as inputs to the RL-based solution. Then, we design a multi-step deep Q-learning algorithm to find good sequential unit commitment decisions. By leveraging the multi-step return, the proposed algorithm explicitly accounts for the fact that the total impacts of a unit commitment decision may not instantly appear in the system operational cost and could influence the costs of many subsequent time steps. Finally, we propose an ensemble framework consisting of a group of deep Q-learning agents which are trained separately in parallel threads with different initial model parameters to final a better UC solution. This design can alleviate the problem that the gradient-based training is prone to being trapped by a locally optimal solution.

The performance of our proposed algorithm, a baseline optimization method, as well as a state-of-the-art RL algorithm \cite{de2021applying} are evaluated on two IEEE test systems. The experimental results show that our proposed algorithm identifies feasible unit commitment solutions with lower costs than both the PPO-based guided tree search algorithm and the MIQP given the same amount of computation time. Moreover, additional scenario analysis demonstrates that our proposed algorithm possesses reasonable generalization capability and can still achieve decent UC results when there is a generation unit or transmission line outage.

The unique contributions of this paper are as follows:

$\bullet$ This paper proposes an optimization method-assisted ensemble multi-step deep reinforcement learning algorithm which synergistically combines the merits of mathematical optimization and reinforcement learning to solve UC problems. 

$\bullet$ The proposed algorithm establishes a candidate action set by solving a series of simplified UC problems, which ensures high-quality unit commitment solutions that satisfy operational constraints.

$\bullet$ The proposed multi-step deep Q-learning algorithm uses a multi-step return to improve the sample efficiency and speed up the process of learning a near-optimal unit commitment solution.

$\bullet$ The ensemble RL framework reduces the operational costs of the power system by mitigating the problem that gradient-based training of neural networks is prone to be trapped by a locally optimal solution. 

$\bullet$ The proposed optimization method-assisted ensemble multi-step deep reinforcement learning algorithm has great generalization capability to solve unit commitment problems when there is a loss of generation units or transmission lines.

The remainder of the paper is organized as follows: Section II gives the formulation of UC problems. Section III introduces the technical methods. Section IV discusses the experimental and algorithm setup as well as the results of numerical studies. Section V makes the conclusion.

\section{Problem Formulation}
In this section, we first give the formulation of Unit Commitment (UC) problems, then we go over the preliminaries of the Markov decision process (MDP). At last, the UC problems are formulated as an MDP.

\subsection{Formulation of Unit Commitment Problems}

As shown in (\ref{eq:uc_objective}), the goal of UC problems is to obtain the optimal schedule of generators which yields the minimum overall operation cost, while subject to certain operational constraints. UC problems for arbitrary optimization horizons can be formulated as follows. Here we take five types of constraints into consideration, specifically, the load/spinning reserve demand constraints (\ref{eq:load})-(\ref{eq:spining}), generation constraints (\ref{eq:generation1})-(\ref{eq:generation2}), ramping limits (\ref{eq:ramp1})-(\ref{eq:ramp3}), minimum up and down time limits (\ref{eq:minimum up1})-(\ref{eq:minimum dn2}), and transmission line capacity constraints (\ref{eq:line constraint}):

\begin{equation}\label{eq:uc_objective}
  \min\sum_{t=1}^T\sum_{i=1}^N \left\{c_i^p(t)+c_i^u(t)+c_i^d(t)\right\},
\end{equation}

\begin{gather}
     s.t.\ \ \ \sum_{i=1}^N p_i(t)=\sum_{j=1}^{M}d_j(t), \ t=1,\cdots,T,  \label{eq:load} \\ 
    \ \ \ \sum_{i=1}^N \Bar{p}_i(t) \geq \sum_{j=1}^{M}d_j(t)+R(t),\ t=1,\cdots,T, \label{eq:spining} \\
        \underline{P}_i v_i(t) \leq p_i(t) \leq \bar{p}_i(t), \ i=1,\cdots,N,  t=1,\cdots,T, \label{eq:generation1} \\
    0 \leq \bar{p}_i(t) \leq \bar{P}_i v_i(t),\ i=1,\cdots,N, t=1,\cdots,T, \label{eq:generation2} \\
    \begin{split} \label{eq:ramp1}
        \bar{p}_i(t) &\leq p_i(t-1) + \text{RU}_iv_i(t-1) +\bar{P}_i(1-v_i(t)) \\
        &+\text{SU}_i[v_i(t)-v_i(t-1)], \ i=1,\cdots,N, t=1,\cdots,T, 
    \end{split} \\
    \begin{split} \label{eq:ramp2}
        \bar{p}_i(t) \leq \bar{P}_iv_i(t+1) +\text{SD}_i[v_i(t)-v_i(t+1)], \\
        i=1,\cdots,N, t=1,\cdots,T, 
    \end{split} \\
    \begin{split} \label{eq:ramp3}
        p_i(t-1) &\leq p_i(t) + \text{RD}_iv_i(t)+\text{SD}_i[v_i(t-1)-v_i(t)]  \\
        & +\bar{P}_i[1-v_i(t-1)],\ i=1,\cdots,N, t=1,\cdots,T, 
    \end{split} \\
        \sum_{t=1}^{\min(G_i,T)}[1-v_i(t)] =0, \ i=1,\cdots,N, t=1,\cdots,T, \label{eq:minimum up1} \\
        \begin{split} \label{eq:minimum up2}
        \sum_{n=t+1}^{\min(t+\text{UT}_i,T)} v_i(n) \geq \sigma_i^u(t)[v_i(t)-v_i(t-1)], \\ 
        \ i=1,\cdots,N,\ t = G_i,\cdots,T-1,
        \end{split} \\
    \sum_{t=1}^{\min(L_i,H)} v_i(t) =0, \ i=1,\cdots,N, \label{eq:minimum dn1} \\
    \begin{split} \label{eq:minimum dn2}
        \sum_{n=t+1}^{\min(t+\text{DT}_i,H)} [1-v_i(n)] \geq \sigma_i^d(t)[v_i(t-1)-v_i(t)], \\ 
        i=1,\cdots,N,\ t = L_i,\cdots,T-1,
    \end{split} \\
    \begin{split} \label{eq:line constraint}
        F_l^- \leq \sum_{i=1}^{N} p_i(t) \Gamma_{i,l}^P -\sum_{j=1}^{M} d_j(t) \Gamma_{j,l}^D \leq F_l^+, \\ l =1,\cdots,L,\ t=1,\cdots,T.
    \end{split}
\end{gather}
where $T$ is the number of time periods of optimization. $N$ and $M$ are the number of units and number of buses. $c_i^p(t),c_i^u(t),c_i^d(t)$ are the production cost, startup cost and shutdown cost of unit $i$ in period $t$, respectively. $p_i(t)$ and $\Bar{p}_i(t)$ are the power output and maximum available output of unit $i$ in period $t$. $d_j(t)$ is the load demand and at bus $j$ in period $t$. $R(t)$ is spinning reserve requirement of the system in period $t$. $\bar{P}_i, \underline{P}_i$ are the capacity and minimum output of unit $i$. $v_i(t)$ is the commitment status of unit $i$ in period $t$. $\text{RU}_i$ and $\text{RD}_i$ are the ramping up and down constraints of unit $i$. $\text{SU}_i$ and $\text{SD}_i$ are the startup and shutdown ramp constraints of unit $i$, $G_i$ and $L_i$ are the numbers of periods during which unit $i$ must be on or off in the beginning. $F_l^-$ and $F_l^+$ are the negative and positive power flow limit of line $j$. $\Gamma_{i,l}^P$ and $\Gamma_{i,l}^P$ are the power transfer distribution factor from unit $i$ to line $l$. $\sigma_i^u(t), \sigma_i^d(t)$ are the number of periods that unit $i$ must be on or off from period $t$ on, which are given in (\ref{eq:mini up})-(\ref{eq:mini dn}) below: 
\begin{gather}
    \sigma_i^u(t)=
    \begin{cases} \label{eq:mini up}
        \min(\text{UT}_i,T-t), \ &\text{if} \ H>G_i, \\
        0, \ &\text{else,}
    \end{cases} \\
    \sigma_i^d(t)=
    \begin{cases} \label{eq:mini dn}
        \min(\text{DT}_i,T-t), \ &\text{if} \ H>L_i, \\
        0, \ &\text{else.}
    \end{cases}
\end{gather}
$\text{UT}_i$ and $\text{DT}_i$ are the minimum up and down time of unit $i$.

The production cost, startup cost, and shutdown cost in (\ref{eq:uc_objective}) are specifically defined as follows:

\subsubsection{Production Cost} Here we use the quadratic production cost function \cite{wood2013power} as in equation (\ref{eq:production_cost}).
\begin{equation}\label{eq:production_cost}
    \begin{split}
        c_i^p(t) =a_i v_i(t)+b_i p_i(t)+c_i p_i^2(t), \\ \forall{i}=1,\cdots,N, \forall{t}=1,\cdots,T,
    \end{split}
\end{equation}
where $a_i, b_i$ and $c_i$ are the coefficients of the quadratic production cost function.
  
\subsubsection{Startup Cost} A mixed-integer linear function for the stair-wise startup cost is formulated in (\ref{eq:start_cost}):
\begin{equation} \label{eq:start_cost}
    \begin{split}
        c_i^u(t) &\geq \text{CU}_i^k \left[ v_i(t)-\sum_{n=1}^k v_i(t-n) \right],  \\
        &\forall{i}=1,\cdots,N, \forall{t}=1,\cdots,T, \forall{k}=1,\cdots,\text{ND}_i,
    \end{split}  
\end{equation}
\begin{equation}
  c_i^u(t) \geq 0, \ \forall{i}=1,\cdots,N, \forall{t}=1,\cdots,T, 
\end{equation}
where $\text{CU}_i^k$ is the stair-wise startup cost of unit $i$ in period $t$. $\text{ND}_i$ is the number of intervals of the staircase startup cost function of unit $i$.

\subsubsection{Shutdown Cost} The shutdown cost is shown in (\ref{eq:shut_cost}):
\begin{equation}\label{eq:shut_cost}
    \begin{split}
        c_i^d(t) \geq \text{CD}_i (v_i(t&-1)-v_i(t)), 
        \\ &\forall{i}=1,\cdots,N, \forall{t}=1,\cdots,T,
    \end{split}
\end{equation}
\begin{equation}
    c_i^d(t) \geq 0, \ \forall{i}=1,\cdots,N, \forall{t}=1,\cdots,T,
\end{equation}
where $\text{CD}_i$ is the shutdown cost of unit $i$.

\subsection{Preliminaries of Markov Decision Process}
 As a classical mathematical framework to formulate sequential decision making problems, Markov Decision Process can be defined as a tuple $(\mathcal{S}, \mathcal{A}, \mathcal{P}, \mathcal{R}, \gamma)$, which consists of  a state space $\mathcal{S}$, an action space $\mathcal{A}$, a state transition probability $\mathcal{P}$, a reward function $\mathcal{R}$ and a discount factor $\gamma$ ($0\leq\gamma\leq1$) \cite{sutton2018reinforcement}. The agent chooses an action $a_t \in \mathcal{A}$ at every time step $t$ depending on the current state $s_t$, and it gains a certain reward $r_{t+1}$, then the environment shifts to the next state $s_{t+1}$ based on $\mathcal{P}(s_{t+1}|s_t,a_t)$. 

The agent aims to find a policy $\pi(a|s)$ that gives the maximum anticipated discounted return $J(\pi)=\mathbf{E}[G(\tau)]$. Here $G(\tau)=\sum_{t=0}^T \gamma^tr_{t+1}$, $T$ is the length of the episode, and $\tau$ is a trajectory of states and actions. In order to demonstrate the value of states and state-action pairs given a policy $\pi$, we give the definition of two crucial value functions $v_\pi(s)$ and $q_\pi(s,a)$:

\begin{align}
v_\pi(s)&={\mathbf{E}_\pi[G_t|S_t=s]}\notag\\
&={\mathbf{E}_\pi\left[\textstyle \sum_{k=0}^T\gamma^k r_{t+k+1}|S_t=s\right]}, \label{eq:v}\\
q_\pi(s,a) &= {\mathbf{E}_\pi[G_t|S_t=s,A_t=a]}\notag\\
&={\mathbf{E}_\pi\left[\textstyle \sum_{k=0}^T\gamma^k r_{t+k+1}|S_t=s,A_t=a\right]}. \label{eq:q}
\end{align}
We define the best policy as $\pi(a|s)=\arg\max_\pi v_\pi(s)$ for all $s \in S$ or $\pi(a|s)=\arg\max_\pi q_\pi(s,a)$ for all $s\in S$ and $a \in A(s)$.

\subsection{Formulate the UC Problems as an MDP}
The UC problems are constructed as an MDP in this subsection. We give the following definition of episode, state, action, and reward functions.

\subsubsection{Episode and Time steps} 
Each operation period is defined as a time step. Since the UC problems are solved daily in the day-ahead market, we formulate them as continuous tasks, which means an episode ends only when no feasible action can be found.

\subsubsection{States}
In order to ensure the environment is Markovian, the state at time $t$ is defined as $s_t = (t, \pmb{v}_t,\pmb{p}_t,\pmb{u}_t,\pmb{d}_t)$, where $t$ is the global time, $\pmb{v}_t$ is a vector of the commitment status $v_i(t)$ of generator $i$ in time $t$ (1 if the unit is on, 0 otherwise), $\pmb{p}_t$ is a vector of the power generation $p_i(t)$ of unit $i$ in time $t$, $\pmb{u}_t$ is a vector of the number of periods that unit $i$ has been running or offline until time $t$, and the transition function of $u_i(t)$ can be formulated as (\ref{eq:u_formulation}):
\begin{equation}\label{eq:u_formulation}
  u_i(t) = 
  \begin{cases}
  u_i(t-1)+1, &\text{if}\ v_i(t) = v_i(t-1), \\
  1,& \text{otherwise.} 
  \end{cases}
\end{equation}
Here $v_i(0)$ is the on/off state of unit $i$ in the beginning of the episode, and $u_i(0)$ is the amount of periods that unit $i$ has been running or offline before the initial period of the episode. Finally, $\pmb{d}_t$ is a vector $[d(t+1),d(t+2),\cdots,d(t+k)]$ of load predictions for the next $k$ periods.

\subsubsection{Actions}
 We define the action $a_t$ at time $t$ as shifting the commitment status of all generators to $\pmb{v}_{t+1}$ in period $t+1$. Because of the operational limits of generators, there are numerous infeasible statuses. To avoid missing the best action, we have to obtain all the feasible actions in the current state. However, the space of the feasible actions set remains prohibitively large even though we can filter out infeasible actions. Besides, it will be extremely difficult for reinforcement learning to learn a good policy from a huge action space. So we use the optimization method to down-select candidate solutions and build the feasible action subset $\mathcal{A}_t$, which will be introduced later in subsection III-A.

\subsubsection{Rewards}
To refrain from the early termination of an episode, the agent receives a large penalty when no feasible action can be found in the current state. So, the reward function is given as (\ref{eq:reward}):
\begin{equation}\label{eq:reward}
    r_{t+1}=-
    \begin{cases}
    C_{t+1},&\text{if}\ \mathcal{A}_{t+1}\neq\varnothing, \\
    \zeta, &\text{if}\ \mathcal{A}_{t+1}=\varnothing.
    \end{cases}
\end{equation} 

\noindent Here $C_{t+1}$ is the negative of the operational cost in period $t+1$ as shown in (\ref{eq:cost}):
\begin{equation}\label{eq:cost} 
    C_{t+1}=\sum_{i=1}^N c_i^p(t+1)+\sum_{i=1}^N c_i^u(t+1)+\sum_{i=1}^N c_i^d(t+1), 
\end{equation}
where the production cost $c_i^p(t+1)$ is derived by solving a one-period economic dispatch (ED) after the commitment status $v_i(t+1)$ is obtained. Here we use $\pmb{u}_t$ to directly calculate the startup cost as in (\ref{eq:reward_startup_cost}):
\begin{equation}\label{eq:reward_startup_cost}
c_i^u(t+1)=
\begin{cases}
\textbf{SCU}_i\left[\min\{\text{ND}_i,u_i(t)\}\right],& \text{if}\ 
v_i(t+1)>v_i(t)\\
0,&\text{otherwise,}
\end{cases}
\end{equation}
where $\textbf{SCU}_i=\left[\text{CU}_i^1,\cdots, \text{CU}_i^{\text{ND}_i}\right]$ is a list of the staircase startup cost of unit $i$, and the symbol $\textbf{SCU}_i[j]$ represents the $j$-th element of the vector $\textbf{SCU}_i$.
  
\section{Technical Methods} 
In this section, we present the proposed optimization method-assisted ensemble RL algorithm. First, we give the process of finding candidate actions using optimization methods, then we introduce a multi-step deep Q-learning algorithm to solve the UC problems. Finally, we design an ensemble framework to further boost performance.

\subsection{Finding Candidate Actions Using Optimization methods}

The process of finding candidate actions can be divided into two parts. First, based on the current state, we solve a tailored UC problem for the next couple of periods to obtain a unit commitment schedule as the cardinal action. Then, given the cardinal action, we obtain more candidate actions by turning on or off more units based on the priority of units. 

\subsubsection{Finding the Base Action} 
Starting from period $t$, the mathematical form of a $H$-period UC problem can be formulated as follows:
\begin{gather}
    \min\sum_{k=t+1}^{t+H}\sum_{i=1}^N \left\{C_i(k)+\omega_{t}\left(v_i(k+1)-v_i(k)\right)\rho_i, \right\} \\
    s.t. \ \ \ (\ref{eq:load})-(\ref{eq:line constraint}) \notag
\end{gather}
where $C_i(k)=c_i^p(k)+c_i^u(k)+c_i^d(k)$ is the production cost of unit $i$ in period $k$. $\omega_{t}$ is a coefficient related to $t$ which is a positive constant when $t>0$ and equals to zero when $t=0$. $\rho_i$ is the average fuel price per output power of unit $i$, which is given in the following formula (\ref{eq:pr}): 
 \begin{equation} \label{eq:pr}
     \rho_i = \frac{a_i+b_i\bar{P}_i+c_i\bar{P}_i^2}{\bar{P}_i}.
 \end{equation}
 
After solving the UC problem above, we can obtain the unit commitment schedule of next $H$-period as $\pmb{v}_{t+1},\cdots,\pmb{v}_{t+H}$ and we set $\pmb{v}_{t+1}$ to be the cardinal action $\pmb{v}^*_{t+1}$of next period.

\subsubsection{Obtaining More Candidate Actions}
Assume there are $X$ units which are turned on or off at period $t+1$ if we take action $\pmb{v}^*_{t+1}$, where $X=\sum_{i=1}^N |v^*_i(t+1)-v_i(t)|$. Then, instead of turning on/off $X$ units, we turn on or off $z$ units which have the higher priority by solving the following single period UC problems:  
\begin{gather}
    \min \sum_{i=1}^N \left( v_i(t+1)-v_i(t) \right)\rho_i, \\
    s.t. \ \ \ v_j(t+1) = v_j(t),\ \forall{j} \in \Theta_t, \\
       \ \ \ \ \sum_{i=1}^N |v_i(t+1)-v_i(t)|=z, \\
       (\ref{eq:load})-(\ref{eq:ramp3}),(\ref{eq:line constraint}) \notag
\end{gather}
where $z$ gradually increases from $\max(X-Y^-,0)$ to $\min(X+Y^+,N)$. $Y^-$ and $Y^+$ denote the parameters of the searching range for unit status change beyond $X$.
$\Theta_{t}$ denotes the set of indexes of the units which cannot be turned on/off due to the minimum up/down-time limit and shutdown ramping limit at period $t+1$.

Note that here we can keep the top $K$ best solutions of the single period UC problem, which means we can obtain $|Z|\times K$ candidate actions, $|Z|$ is the number of elements in the range of $z$. Finally, there are $|Z|\times K+1$ candidate actions in the action subset $\mathcal{A}_t$.

\subsection{Multi-Step Deep Q-Learning for UC Problems}
For the purpose of solving MDPs with continuous state space \cite{mnih2013playing}, Deep Q-learning integrates the standard Q-learning with a deep neural network named deep Q network (DQN) $Q(s_t,a_t|\theta)$ to estimate the action-value function in (\ref{eq:q}). We use Adam gradient descent to train DQN to minimize the mean-squared temporal difference error $L(\theta)$:
\begin{equation}\label{eq:Q_loss}
    L(\theta)=\mathbf{E}_{(s,a,r,s')\sim \mathcal{D}}\left(r+\gamma\max_{a'} Q(s',a'|\theta')-Q(s,a|\theta)\right)^2
\end{equation}
where $Q(s',a'|\theta')$ is the target neural network with the same structure as $Q(s_t,a_t|\theta)$. In order to make the training process stable, we update the parameters $\theta'$ of the target network from the parameters $\theta$ of $Q(s_t,a_t|\theta)$ in a periodical manner. $\mathcal{D}$ is the replay buffer to collect the transition tuples $(s,a,r,s')$.

However, adopting DQN to solve UC problems without modifications can be inefficient, since taking an action may not only affect the next reward, but also influence the rewards multiple steps later. We need several updates to propagate the reward to the related preceding states and actions \cite{mnih2016asynchronous}, which makes the training process both time-consuming and tremendously sample-inefficient.

To address this issue, we use the multi-step return method \cite{watkins1989learning} to update the action-value function $Q(s_t,a_t|\theta)$: 
\begin{equation}
    L(\theta) = (Q(s_t, a_t|\theta) - R(t))^2,
\end{equation}
where $R(t)$ is defined as:
\begin{equation}\label{eq:n_step_return}
    R(t) = r_{t+1}+\gamma r_{t+2}+\cdots+\gamma^{n-1}r_{t+n}+\max_{a' \in \mathcal{A'}} \gamma^n Q(s_{t+n},a'|\theta')
\end{equation}
Note that we make the agent-environment interact for $n$ steps to acquire the rewards $r_{t+k}, k = 1,\dots, n$ and the $n$-step next state $s_{t+n}$, and then calculate (\ref{eq:n_step_return}). 

Using the $n$-step return, the long-term as well as the short-term impacts of taking an action can be studied by propagating toward the exact reward, instead of bootstrapping from the target network $Q(s,a|\theta')$. Therefore, the learning efficiency of UC problems is prominently enhanced by applying multi-step deep Q-learning algorithm \cite{qin2021solving}.

\subsection{Ensemble RL Framework for UC Problems}

We now present a multi-threaded ensemble reinforcement learning framework. The aim of designing this framework is to mitigate the problem of reaching a bad local optimal solution from a single random initialization of deep Q-network's parameters. Specifically, we initialize the aforementioned multi-step deep Q-learning algorithm in different threads with different random seeds and run them in parallel. The pseudocode for training the ensemble multi-step deep Q-learning for UC problems is shown in algorithm \ref{alg1}. After the ensemble multi-step deep Q-learning algorithm is trained, the $M$ instances of RL agents can be run in parallel during the testing phase. The multi-step deep Q-learning agent that identifies the unit commitment solution with the lowest operational cost will be selected as the final solution.


\begin{algorithm}[htbp]
	\caption{Training of Ensemble Multi-Step Deep Q-learning for UC Problems}
	\label{algo_multiq}
	Initialize $M$ evaluation $Q$-network with random parameters $\theta_1\cdots\theta_M$ \\
	Initialize $M$ target $Q$-network with parameters $\theta'_1=\theta_1,\cdots,\\
	\theta'_M=\theta_M$ 
	\begin{algorithmic}[1]
	\For{$\text{thread}\ m=1,\cdots,M$}{}
	\State Input historical data set of $N_d$ days and set day $d=1$ 
	\State Initialize replay buffer $\mathcal{D}$ as a queue with a maximum 
	\State length of $n$ 
	\State Initialize learning counter $\nu=0$ 
	\For{$\text{episode}=1,\cdots,\Gamma$}{}
	\State Input historical load data of day $d$
	\State Formulate initial state $s_1$ of day $d$
	
	\For{$t=1,\cdots,T$}{}
	\State Obtain candidate action set $\mathcal{A}_t$ of state $s_t$ using 
	\Statex[3] optimization method.
	\State With $\epsilon$ choose a random action $a_t$ from $\mathcal{A}_t$,
	\Statex[3] otherwise choose $a_t=\text{max}_{a\in \mathcal{A}_t}Q(s_t,a|\theta'_m)$.
	\State Obtain the schedule of units on next period $t+1$  
	\Statex[3] based on action $a_t$.
	\State Solve a single period ED and calculate reward 
	\Statex[3] $r_{t+1}$ according to (\ref{eq:reward}).
	\State Calculate $\pmb{u}_{t+1}$ according to (\ref{eq:u_formulation}) and then
	\Statex[3] formulate the next state $s_{t+1}$.
	\State Use optimization method to calculate $\mathcal{A}_{t+1}$.
    \State Set $\varepsilon_t=1 \ \text{if}\ \mathcal{A}_{t+1}=\varnothing \ \text{else} \ 0$
	\State Store $(s_t,a_t,r_{t+1},s_{t+1},\mathcal{A}_{t+1},\varepsilon_t)$ in $\mathcal{D}$
	\If{$\text{length}(\mathcal{D})=n$ or $\varepsilon_t = 1$}{}
	\State $R=0 \ \text{if}\ \varepsilon_t=1 \ \text{else} \ \max_a Q(s_{t+1},a|\theta'_m)$
	\For{$i=t,t-1,\cdots,t-\text{length}(\mathcal{D})$,}{}
	\State Set $R= r_i+\gamma R$
	\State Perform a gradient descent step on
	\Statex[5] $(R-Q(s_i,a_i|\theta_m))^2$
	\EndFor
	\State Set $\nu = \nu+1$
	\If{$\text{mod}(\nu,I_{target})=0$}{}
	\State Update $\theta'_m=\theta_m$
	\EndIf
	\EndIf
	\EndFor
	\If{day $d$ is over}{}
	\State $d=\mathrm{mod}(d+1,N_d)$
	\EndIf
	\EndFor
	\EndFor
	\end{algorithmic}
	\label{alg1}
\end{algorithm}
 
Here are some key implementation details of algorithm \ref{alg1}.

$\bullet$ Q-network structure: We adopt the feed-forward neural networks as the Q-networks, whose inputs are state-action pairs and outputs are the resulting Q value. The reason for selecting this architecture is that it can scale linearly with the number of generators.

$\bullet$ Episode initialization: Since UC problems are formulated as continuous tasks, which means we aim to maximize the overall reward received in all training episodes, we get the initial state of the current day from the final period of the previous day. Thus, the historical data of the next day will not be utilized for training until a policy that meets all load demands of the current day is found.

$\bullet$ Global Time encoding: In order to present the periodic nature of the problem, the global time step $t$ is decomposed into two coordinates $[cos(2\pi t/24),sin(2\pi t/24)]$ \cite{gao2021consensus}, which varies from 0 to 23.

\section{Numerical Studies}
\subsection{Experimental and Algorithm Setup}
In this subsection, we give the experimental and algorithm setups. All algorithms are executed on a server with a 32-core AMD Ryzen Threadripper 3970X 3.7GHz CPU. 

\subsubsection{Experimental Setup} 
We apply the proposed method to solve a 48-hour period UC problem for the IEEE 118-bus system and 300-bus system \cite{zimmerman2010matpower}. The parameters of units such as minimum up and down time limit are obtained from \cite{carrion2006computationally}. The detailed experimental setup for two systems can be found in our open-source repository\footnote{\url{https://github.com/jqin020/Emsemble-Deep-RL-for-UC-problems}}. The aforementioned minimum and maximum of staircase startup cost $\textbf{CU}$ of all units are equivalent to their hot start cost and cold start cost. Initial on/off time is the number of periods that one generator has been running or offline before the first period of the starting day. Here we give four different initial status setups to verify the generalization ability of our algorithm. The historical load data of the California Independent System Operator (CASIO) \cite{CASIO} from January 1, 2021 to July 5, 2021 is used and scaled to be suitable for the two IEEE systems. Note that we use 90 days for training, one week for validation, and two weeks from different months for testing.

\subsubsection{RL Algorithm Setup}

The hyperparameters of the benchmark PPO guided tree search \cite{de2021applying} and the proposed ensemble multi-step deep Q-learning algorithm are summarized in Table \ref{table:Algorithm Setup}. We tune all parameters separately to achieve the optimal performance. 
\begin{table}[h]
	\caption{Hyperparameters of Benchmark and Proposed RL Algorithms}
	\begin{tabular}{L{3.2cm} L{3.0cm} L{1.3cm}}
		\toprule  
		\multirow{10}{*}{Ensemble multi-step deep-Q} &Number of threads $M$ &10  \\ 
		    & Number of steps $n$  &24  \\ 
		    & Load forecast steps $k$ &9 \\
		    & Learning rate $\alpha$ &0.0001 \\
		    & Update frequency $I_{target}$ &60 \\
		    & Greedy Arrange $\bar{\epsilon}$ &[0.01,1.0]\\
		    & Optimization Horizon $H$ &2 \\ 
		    & Search Down $Y^-$ &1 \\
		    & Search Up $Y^+$ &1 \\
		    & Top $K$ Best Actions  &1 \\ \hline    
		\multirow{6}{*}{PPO guided tree search} & Actor learning rate &0.003\\ 
		    & Critic learning rate &0.001\\ 
		    & Number of epochs & $80$ \\
		    & Clipping $\xi$ & $0.2$ \\
		    & Search Depth $H$ &2 \\
		    & Branching Threshold $\rho$ &0.1 \\
		    \hline        
        \multirow{5}{*}{Shared parameters} & Number of hidden units & $\{150,150\}$ \\
            & Number of hidden layers &1 \\
            & Discount factor $\gamma$ &0.99\\ 
            & Number of episode &50\\ 
            & Optimizer &Adam \\       
		\bottomrule
	\end{tabular}
	\label{table:Algorithm Setup}
\end{table}

\subsection{Performance Comparison}
\begin{figure}[htbp]     
    \centering
    \subcaptionbox{118-bus system}{\includegraphics[width=6.5cm]{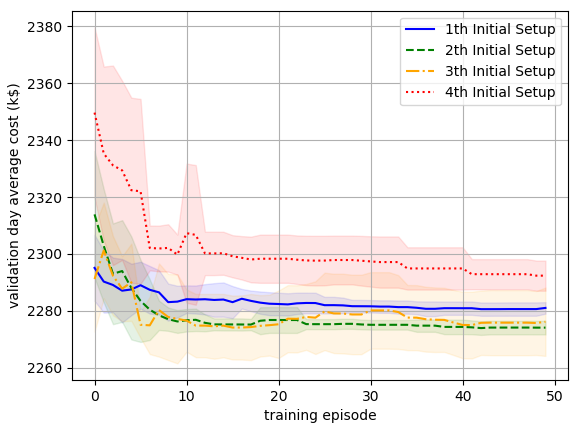}}
    \subcaptionbox{300-bus system}{\includegraphics[width=6.5cm]{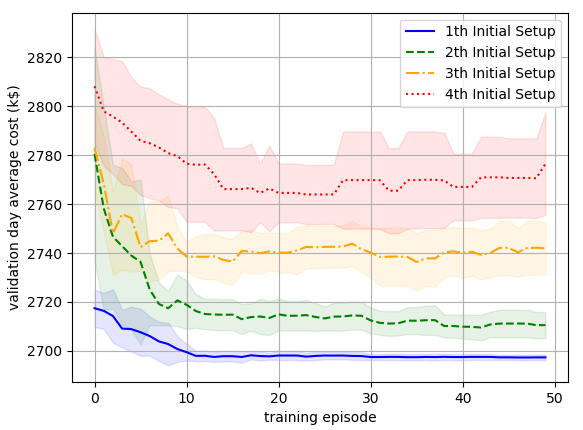}}
    \caption{The average daily cost of validation days}
    \label{fig2}
\end{figure}

In this subsection, we compare the performance of the PPO guided tree search, the MIQP algorithm with Gurobi 9.1 \cite{gurobi} solver, and our proposed algorithm. We set the time limit to 10 minutes and the MIP gap to 0.1\% for Gurobi in all following analyses. Note that the optimization periods of MIQP are 48 hours and we obtain the operation cost of the first day from the optimization result. The mean daily operational cost of our proposed algorithm during the validation days of two IEEE systems under four initial status setups are reported after every training episode in Fig. \ref{fig2}. The beginning commitment status of generators on the first validation day is the same during the training process. The solid curves and shaded areas represent the average values and standard deviations across different runs in 10 threads, respectively.

As shown in Fig. \ref{fig2}, the average daily costs of validation days decrease rapidly as the training continues and maintain a low level under all four initial status setups. When training processes are done, the testing days are adopted to evaluate the algorithms. For PPO guided tree search and our proposed algorithm, we use the network parameters that yield the minimum average daily costs of the validation dataset for testing. The average daily operation cost of testing days across four initial status setups of the two IEEE systems are summarized in Table \ref{tab2} and Table \ref{tab3}, respectively. The percentage variation of both the PPO guided tree search and the proposed Ensemble N-step Q-learning from the MIQP, $\delta_1$ and $\delta_2$ respectively, are also given to compare their performance. 

\begin{table}[htbp]
\caption{Daily Operation Cost of the 118-bus System}
\begin{center}
\begin{tabular}{c|c|c|c|c|c}
\hline
\thead{Week\\-Day} &\thead{PPO tree \\  search $(\$)$} &\thead{Ensemble \\N-step Q $(\$)$} &\thead{MIQP\\ $(\$)$} &$\delta_1 (\%)$ &$\delta_2 (\%)$ \\
\hline
1-1	&2,816,927 	&2,251,095 	&2,245,754 	&1.45	&0.24\\
\hline
1-2	&2,647,468  	&2,106,565      &2,073,649 	   &3.30	&1.59\\
\hline
1-3	&2,853,497 	&2,322,511 	&2,308,647 	&4.82	&0.60\\
\hline
1-4	&2,870,898 	&2,344,690	&2,327,352 	&5.64	&0.74\\
\hline
1-5	&2,876,388 	&2,343,923 	&2,323,781 	&5.93	&0.87\\
\hline
1-6	&2,847,783 	&2,339,326	&2,301,149 	&5.95	&1.66\\
\hline
1-7	&2,831,246 	&2,330,793 	&2,285,692 	&5.81	&1.97\\
\hline
2-1	&3,058,294 	&2,535,563 	&2,511,276 	&0.91	&0.97\\
\hline
2-2	&3,021,903 	&2,524,269     &2,490,130	   &2.64	&1.37\\
\hline
2-3	&3,010,909 	&2,524,673 	&2,487,975 	&3.90	&1.48\\
\hline
2-4	&2,906,699 	&2,399,318 	&2,348,203 	&3.93	&2.18\\
\hline
2-5	&2,885,750 	&2,412,025 	&2,365,566 	&4.03	&1.96\\
\hline
2-6	&3,081,577 	&2,647,766	&2,606,349	&2.87	&1.59\\
\hline
2-7	&2,796,340 	&2,935,628 	&2,907,972 	&2.01	&0.95\\
\hline
\end{tabular}
\label{tab2}
\end{center}
\end{table}

\begin{table}[htbp]
\caption{Daily Operation Cost of the 300-bus System}
\begin{center}
\begin{tabular}{c|c|c|c|c|c}
\hline
\thead{Week\\-Day} &\thead{PPO guided \\ tree search $(\$)$} &\thead{Ensemble \\N-step Q $(\$)$} &\thead{MIQP\\ $(\$)$} &$\delta_1 (\%)$ &$\delta_2 (\%)$ \\
\hline
1-1	&2,844,044 	&2,736,283 	&2,702,568 	&5.23	&1.25\\
\hline
1-2	&2,548,734 	&2,503,574  &2,484,569	 &2.58	&0.76\\
\hline
1-3	&2,881,019 	&2,769,865 	&2,755,564 	&4.55	&0.52\\
\hline
1-4	&2,894,951 	&2,798,182 	&2,780,654 	&4.11	&0.63\\
\hline
1-5	&2,934,550 	&2,798,752 	&2,778,608 	&5.61	&0.72\\
\hline
1-6	&2,902,320 	&2,772,518	&2,746,544 	&5.67	&0.95\\
\hline
1-7	&2,866,755 	&2,761,019 	&2,738,949 	&4.67	&0.81\\
\hline
2-1	&3,124,671 	&3,069,440 	&3,011,760 	&3.75	&1.92\\
\hline
2-2	&3,046,178 	&3,003,487  &2,977,668	&2.30	&0.87\\
\hline
2-3	&3,054,133 	&3,003,764 	&2,972,566 	&2.74	&1.05\\
\hline
2-4	&2,890,670 	&2,852,659 	&2,807,185 	&2.97	&1.62\\
\hline
2-5	&2,908,886 	&2,865,494 	&2,831,723 	&2.72	&1.19\\
\hline
2-6	&3,193,841 	&3,152,416	&3,121,763 	&2.31	&0.98\\
\hline
2-7	&3,552,765 	&3,499,896	&3,471,958 	&2.33	&0.80\\
\hline
\end{tabular}
\label{tab3}
\end{center}
\end{table}

From Table \ref{tab2} and Table \ref{tab3} we can see the average daily operation cost, as well as percent variation of the proposed algorithm from MIQP is much smaller than that of the PPO guided tree search. Additionally, the total computation time of testing weeks of two systems are shown in Table \ref{tab4}. Here we only compare the testing time of RL-based and MIQP algorithms since the training process of RL-based algorithms can be done in an offline manner.

\begin{table}[htbp]
    \centering
    \caption{Total Computation Time of Testing weeks}
    \begin{tabular}{c c c c c} 
        \hline
        \toprule  
        & &\thead{PPO tree \\  search $(s)$} &\thead{Ensemble \\N-step Q $(s)$} &\thead{MIQP\\ $(s)$} \\
        \hline
        \multirow{2}{*}{118-bus} &First test week &583 &113 &4,278 \\
        &Second test week &601 &122 &4,283 \\ \hline
        \multirow{2}{*}{300-bus} &First test week &741 &125 &4,355 \\
        &Second test week &756 &144 &4,353 \\
        \bottomrule
    \end{tabular}
    \label{tab4}
\end{table}

\begin{figure}[htbp]     
    \centering
    \subcaptionbox{118-bus system}{\includegraphics[width=6.5cm]{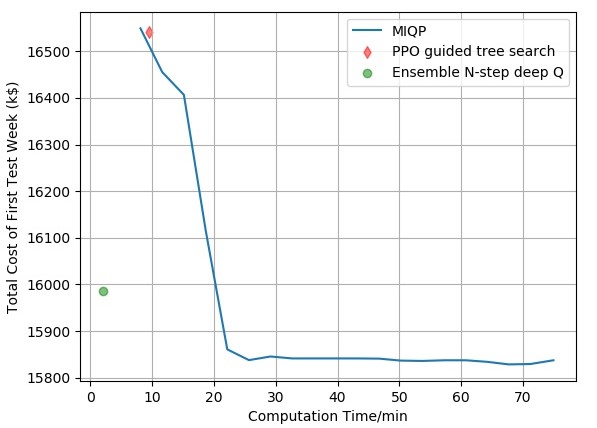}}
    \subcaptionbox{300-bus system}{\includegraphics[width=6.5cm]{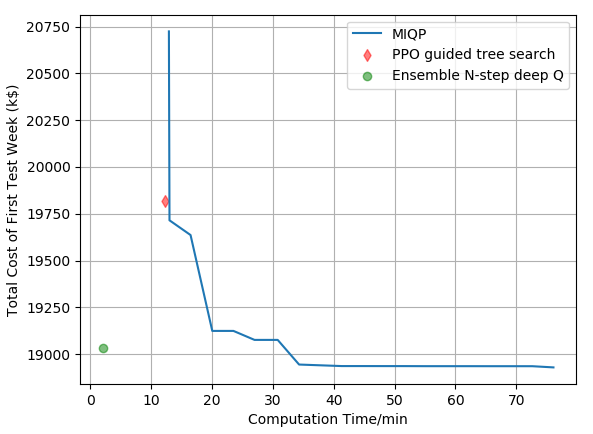}}
    \caption{Comparison of three algorithms}
    \label{fig3}
\end{figure}

To further demonstrate the improvement of our proposed algorithm, the total operation cost of the first test week of two systems computed by PPO guided tree search, ensemble multi-step deep Q-learning, and MIQP with respect to the computation time are shown in Fig. \ref{fig3}. From the figure, we can see that the performance of PPO guided tree search is close to MIQP while our proposed algorithm substantially surpasses MIQP. In other words, to identify a unit commitment solution of the same total operation cost, our proposed ensemble n-step deep Q-learning only needs a fraction of the computation time required by PPO-guided tree search and the MIQP algorithm. Given sufficient computation time, the MIQP algorithm will as expected eventually identify a solution, which has a lower operational cost than our proposed method.

\subsection{Ablation Study}

\begin{figure}[htbp]     
    \centering
    \subcaptionbox{118-bus system}{\includegraphics[width=6.5cm]{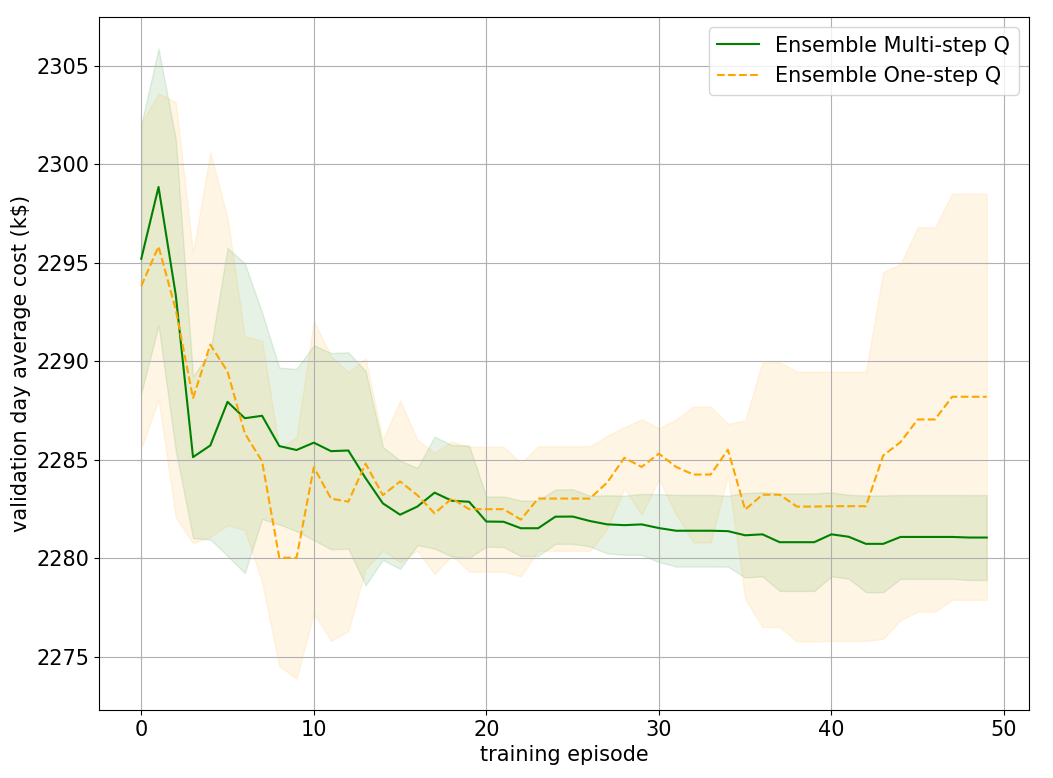}}
    \subcaptionbox{300-bus system}{\includegraphics[width=6.5cm]{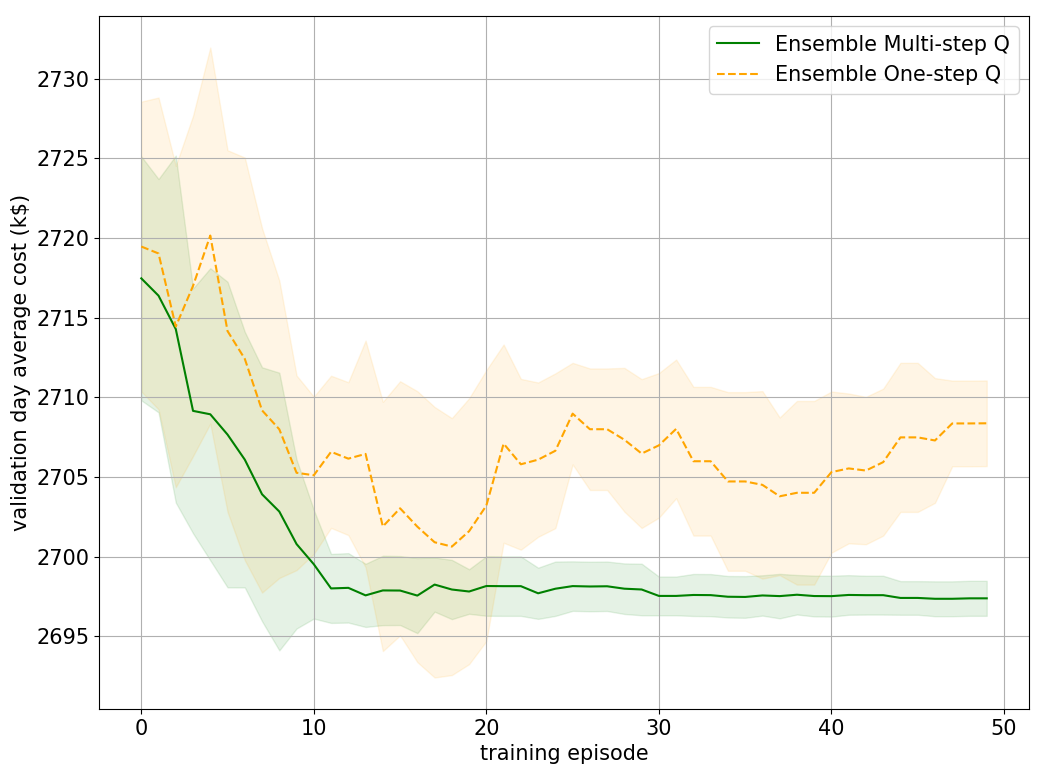}}
    \caption{Comparison of one-step and multi-step return}
    \label{fig4}
\end{figure}

In this subsection, we study the impact of systematically removing some features on our proposed algorithm. We start by comparing the performance of using one-step return and using multi-step return during the training process under the first initial status setup. As shown in Fig. \ref{fig4}, the average daily costs of validation days calculated by ensemble multi-step deep Q-learning stabilize at a lower level than that of ensemble one-step deep Q-learning, and the standard deviations of average daily costs across different runs received by using multi-step return are much smaller than that of using one-step return for both systems.

After the training process ends, we summarize the average total cost of two test weeks under the third initial status setup using different combinations in Table \ref{tab5}. Note that here we use the final parameters of the neural networks for both ensemble one-step and multi-step deep Q-learning, and we calculate the average total cost of $M$ runs for one-step and multi-step deep-Q learning. It can be seen that the total costs of test weeks are smaller when using a multi-step return and an ensemble framework.

\begin{table}[htbp]
    \centering
    \caption{Total cost of Test Weeks Using Different RL Techniques}
    \begin{tabular}{c c c c} 
        \hline
        \toprule  
         & &Traditional $(k\$)$ &Ensemble $(k\$)$ \\
        \hline
        \multirow{2}{*}{118-bus} &One-step return &34,373 &34,197  \\
        &Multi-step return &34,152 &33,985 \\ \hline
        \multirow{2}{*}{300-bus} &One-step return &41,087 &40,703  \\
        &Multi-step return &40,719 &40,488\\
        \bottomrule
    \end{tabular}
    \label{tab5}
\end{table}

\subsection{Generalization Capability} 

In this subsection, we show the generalization capability of our proposed algorithm under some unforeseen operating conditions, i.e. losing one generation unit or one transmission line due to regular maintenance. First, we show the generation ability of ensemble multi-step deep-Q learning as well as PPO guided tree search when losing one unit. Note that for our proposed algorithm and PPO guided tree search, we set the on/off status and power output of the losing unit to zero and set the time that the losing unit has been on/off to its minimum on/off time limit. Table \ref{tab6} gives the total costs of testing weeks of three algorithms for two systems when losing one unit and we can see that our proposed algorithm yields smaller total operation costs than PPO guided tree search, which are very close to MIQP for both systems.  

\begin{table}[htbp]
    \centering
    \caption{Total Cost of Testing weeks when Losing One Unit}
    \begin{tabular}{c c c c c} 
        \hline
        \toprule  
        &Loss of Unit &\thead{PPO tree \\  search $(k\$)$} &\thead{Ensemble \\N-step Q $(k\$)$} &\thead{MIQP\\ $(k\$)$} \\
        \hline
        \multirow{6}{*}{118-bus} &Unit 3 &34,699 &33,884 &33,514 \\
        &Unit 16 &34,762  &33,916 &33,519 \\ 
        &Unit 28 &34,786 &33,867 &33,505 \\
        &Unit 35 &34,725 &33,886 &33,513 \\
        &Unit 47 &34,687 &33,880 &33,511 \\
        &Avg. Time &659$s$ &128$s$ &4,273$s$ \\\hline
        \multirow{7}{*}{300-bus} &Unit 5 &41,737 &40,368   &40,067  \\
        &Unit 14 &41,535  &40,263   &40,071  \\
        &Unit 26 &41,746  &40,412   &40,069  \\
        &Unit 37 &41,663  &40,316   &40,074  \\
        &Unit 43 &41,783  &40,365   &40,070  \\
        &Unit 58 &41,632  &40,309   &40,074  \\ 
        &Avg. Time &804$s$ &146$s$ &4,357$s$ \\
        \bottomrule
    \end{tabular}
    \label{tab6}
\end{table}

Then, the total operation costs of testing weeks of three algorithms for two systems when losing one transmission line are given in Table \ref{tab7}. Similar to the scenario where systems lose one unit, our proposed algorithm shows greater generalization capability than the PPO guide tree search when losing one line. The baseline RL algorithm may not be able to generalize to unforeseen system operation states. By leveraging a simplified optimization method to identify candidate solutions and combining it with the RL algorithm, the generalization capability of the proposed method is enhanced significantly.

\begin{table}[htbp]
    \centering
    \caption{Total Cost of Testing weeks when Losing One Line}
    \begin{tabular}{c c c c c} 
        \hline
        \toprule  
        &Loss of Line &\thead{PPO tree \\ search $(k\$)$} &\thead{Ensemble \\N-step Q $(k\$)$} &\thead{MIQP\\ $(k\$)$} \\
        \hline
        \multirow{7}{*}{\centering 118-bus} &Line 3 &34,843 &34,225 &33,726 \\
        &Line 38 &34,742 &34,376 &33,779 \\ 
        &Line 61 &34,667 &33,948 &33,505 \\ 
        &Line 98 &34,761 &33,852 &33,497 \\
        &Line 139 &35,010 &34,155 &33,781 \\
        &Line 152 &34,688 &33,880 &33,508 \\ 
        &Avg. Time &594$s$ &128$s$ &4,273$s$ \\ \hline
        \multirow{7}{*}{300-bus} &Line 10 &41,816 &40,301 &40,070 \\
        &Line 42 &41,604 &40,508 &40,216 \\ 
        &Line 60 &41,650 &40,639 &40,304 \\ 
        &Line 92 &41,872 &40,239 &40,049 \\
        &Line 224 &41,549 &40,368 &40,073 \\
        &Line 337 &41,824 &40,266 &40,115 \\ 
        &Avg. Time &745$s$ &134$s$ &4,354$s$ \\
        \bottomrule
    \end{tabular}
    \label{tab7}
\end{table}

\section{Conclusion}
This paper proposes an optimization method-assisted ensemble deep reinforcement learning algorithm to solve unit commitment problems. We establish a candidate action set by solving simplified optimization problems. Multi-step return is used to speed up the learning process and improve the sample efficiency of the reinforcement learning agent. The proposed ensemble framework can mitigate the adverse effects that the gradient-based training could lead to a bad local optimal solution.
Numerical studies show that given a time limit of solution, our algorithm can achieve a better performance than the benchmark PPO guided tree search algorithm as well as MIQP. Furthermore, our proposed optimization method-assisted ensemble deep reinforcement learning algorithm has great generalization ability under unforeseen operating conditions. In the future, we plan to further improve the scalability of the proposed algorithm and tackle the security-constrained unit commitment problems on larger power systems.


 

\bibliographystyle{IEEEtran}
\bibliography{ref}

\newpage

 





\end{document}